\documentstyle[14pt]{article}
\begin{document}
\begin{center}
{\bf JAYNES-CUMMINGS MODEL WITH DEGENERATE ATOMIC LEVELS\/}\\ [1cm]
{\bf V.~A.~Reshetov\/}\\ [1cm]
{\it Department of Physics, Tolyatti Pedagogical Institute,
13 Boulevard Korolyova, 445859 Tolyatti, Russia\/}\\[1cm]
\end{center}
\begin{abstract}
     The Jaynes-Cummings model describing the interaction of a
single linearly-polarized  mode of the quantized electromagnetic
field with an isolated two-level atom is  generalized  to  the
case of  atomic  levels  degenerate  in the projections of the
angular momenta on the quantization axis,  which  is  a  usual
case in the experiments. This generalization,
like the original model, obtains the explicit solution.
     The model  is  applied to calculate the dependence
of the atomic level  populations  on  the  angle  between  the
polarization of cavity field mode and that of the laser
excitation pulse in the experiment  with  one-atom  micromaser.\\[5mm]
\end{abstract}

     The Jaynes-Cummings model [1] describes the interaction of a
single linearly-polarized  mode of the quantized electromagnetic
field with  an  isolated  two-level  atom.  The full set of states
of the system atom+field is
     $$ \vert        n,\alpha>=\vert       n>\cdot       \vert
\alpha>,~~~,n=0,1,...,~~~\alpha=b,c,~$$
     where~$n$~is the  number  of  photons  in the field mode,
while~$b$~and~$c$~denote the upper  and  lower  atomic  levels
correspondingly. This model is applied successfully to analyse
the results of the experiments with one-atom micromasers (see,
e.g., [2]).  However,  the  levels  of  an  isolated  atom are
degenerate in the projections of the total  elctronic  angular
momenta on the quantization axis, so that the original
Jaynes-Cummings model becomes, in general, invalid.

     Now, let us take into account  the  degeneracy  of  atomic
levels. Then, the full set of states of the system becomes
     $$ \vert n,J_{\alpha},m_{\alpha}>=\vert n>\cdot
\vert J_{\alpha},m_{\alpha}>,~n=0,1,...,~
m_{\alpha}=-J_{\alpha},...,J_{\alpha},~\alpha=b,c,~$$
     where~$J_{b}$~and~$J_{c}$~are the  values  of  the  total
electronic angular     momenta     of     resonant     levels,
while~$m_{b}$~and~$m_{c}$~are their    projections    on   the
quantization axis - the cartesian axis Z,  which  is  directed
along the polarization vector of the field mode.

     The Hamiltonian of the system may be written as
     \begin{equation}
     {\hat H}={\hat H_{F}}+{\hat H_{A}}+{\hat V}~,
     \end{equation}
     where
     $${\hat H_{F}}=\hbar \omega {\hat a}^{+}{\hat a}$$
     is a free-field Hamiltonian,
     $${\hat H_{A}}={1 \over 2} \hbar \omega_{0}
       ( {\hat n}_{b}-{\hat n}_{c} )$$
     is a free-atom Hamiltonian,
     $${\hat V}=-({\hat {\bf D\/}}{\hat {\bf E\/}})$$
     is an operator of field-atom interaction, while
~${\hat a}^{+}$~and~${\hat  a}$~are  the  operators   of   the
creation and   annihilation  of  photons  with  the  frequency
~$\omega$~in the field mode,
  $${\hat n}_{\alpha}=\sum_{m_{\alpha}=-J_{\alpha}}^{J_{\alpha}}
   \vert J_{\alpha},m_{\alpha}><J_{\alpha},m_{\alpha}\vert,
  ~\alpha=b,c,~$$
     are the operators of total populations of resonant atomic
levels~$b$~and~$c,$~~$\omega_{0}$~is the   frequency   of  the
optically-allowed atomic transition~${J_{b} \to J_{c}},$~
     $${\hat {\bf  E\/}}={\bf e\/} {\hat a}+{\bf e\/}^{*}{\hat
a}^{+},~$$
   $$ {\bf e\/}=\imath {\bf l\/}_{z}
\sqrt{ {2 \pi \hbar \omega \over V}},$$
     is the electric field intensity operator,~$V$~and~${\bf l
\/}_{z}$~being the resonator cavity volume and the unit vector
of the cartesian axis Z,
     $${\hat {\bf D\/}}={\hat {\bf d\/}}+{\hat  {\bf d\/}}^{+},$$
  $${\hat {\bf d\/}}=\sum_{m_{b},m_{c}}
{\bf d\/}^{J_{c}J_{b}}_{m_{c}m_{b}}\cdot
   \vert J_{c},m_{c}><J_{b},m_{b}\vert,$$
     is the dipole moment operator of the atomic transition
     ~${J_{b} \to J_{c}},$~which matrix elements  are  defined
through Wigner 3j-symbols (see, e.g., [3]):
     $$(d_{q})^{J_{b}J_{c}}_{m_{c}m_{b}}=d(-1)^{J_{b}-m_{b}}
     \left(\matrix{J_{b}&1&J_{c} \cr -m_{b}&q&m_{c}}\right),$$
    ~$d=d(J_{b}J_{c})$~-being a reduced matrix element and~$d_
{q}~(q=-1,0,1)$~- are the circular components of vector~${\bf d
\/}.$~

     In the interaction representation
     $${\hat f}_{I}=\exp{\left({\imath {\hat H}_{0} t \over \hbar}
\right)}
\cdot{\hat f}   \cdot   \exp{\left(-{\imath  {\hat  H}_{0}  t  \over
\hbar}\right)},$$
     where
     $${\hat H}_{0}=\hbar \omega \left\{ {\hat a}^{+}{\hat a}+
     {1 \over 2}({\hat n}_{b}-{\hat n}_{c})\right\},$$
     the operators~${\hat  a}$~and~${\hat  {\bf  d\/}}$~obtain
the oscillating factors
     $${\hat  a}_{I}={\hat  a}\cdot\exp(-\imath \omega t),~~~
     {\hat {\bf  d\/}}_{I}={\hat {\bf d\/}}\cdot
\exp(-\imath \omega t).$$
     Then, in the rotating wave approximation, when  the  terms
oscillating with   double   frequences   are   neglected,  the
Hamiltonian (1) becomes
       $${\hat H}_{I}={\hat H}_{0}-\hbar {\hat \Omega},$$
     where
     $${\hat \Omega}={\delta \over 2}({\hat n}_{b}-{\hat n}_{c})
     +\imath g ({\hat a}{\hat p}^{+}-{\hat a}^{+}{\hat p}),$$
     while
     $$\delta = (\omega - \omega_{0})$$
     is the frequency detuning,
     $$g=\sqrt{{2\pi d^{2}\omega \over \hbar V}}$$
     and
     $${\hat p}=\sum_{m}\alpha_{m}\cdot \vert J_{c},m><J_{b},m
\vert~,$$
     $$\alpha_{m}=(-1)^{J_{b}-m}
     \left(\matrix{J_{b}&1&J_{c} \cr -m&0&m}\right).$$

     From the equation
     $${d {\hat \sigma} \over dt} = {\imath \over \hbar}
\left[ {\hat \sigma},{\hat H}\right]$$
     for the system density matrix~${\hat \sigma}$~follows the
equation
     \begin{equation}
     {d {\hat \rho} \over dt} = \imath
\left[ {\hat \Omega},{\hat \rho}\right]
     \end{equation}
     for the density matrix
 $${\hat \rho} = \exp{\left({\imath {\hat H}_{0} t \over \hbar}
\right)} \cdot{\hat \sigma}   \cdot
\exp{\left(-{\imath  {\hat  H}_{0}  t  \over \hbar}\right)}$$
     in the interaction representation. The formal solution of
the equation (2) is obtained immediately
 $${\hat \rho} = \exp{\left(\imath {\hat \Omega} t \right)}
\cdot{\hat \rho}_{0}   \cdot
\exp{\left(-\imath  {\hat  \Omega} t\right)},$$
     where~${\hat \rho}_{0}$~is  the initial density matrix of
the system.

     In order to obtain the average value
     $$<{\hat f}> = Tr\left({\hat \rho}{\hat f}_{I}\right)$$
     of any operator~${\hat f}$~it is  necessary  to  calculate
the matrix elements
     $$<n,J_{\alpha},m \vert\exp{(\imath {\hat \Omega}t)}
\vert n_{1},J_{\beta},m_{1}>,~\alpha,\beta=b,c,$$
     of the  evolution  operator.  The   explicit   analytical
expressions for  these matrix elements may be derived with the
use expansion
     $$\exp{(\imath {\hat \Omega}t)}=\sum_{n=0}^{\infty}
     {(\imath {\hat \Omega}t)^{n} \over n!}~,$$
     since the operator
     $${\hat \Omega}^{2} = {\delta^{2} \over 4} {\hat 1} +
g^{2}\cdot \left\{{\hat R}_{c}{\hat n}+{\hat R}_{b}({\hat n}+1)
     \right\},$$
     where
  $${\hat R}_{\beta}=\sum_{m}\alpha_{m}^{2}\cdot
   \vert J_{\beta},m><J_{\beta},m\vert,
  ~\beta=b,c,~$$
     is diagonal:
     $${\hat \Omega}^{2}\vert n,J_{b},m>=
      \Omega^{2}_{n+1,m}\vert n,J_{b},m>,$$
     $${\hat \Omega}^{2}\vert n,J_{c},m>=
      \Omega^{2}_{n,m}\vert n,J_{c},m>.$$
     Here
     \begin{equation}
      \Omega_{n,m}=\sqrt{{\delta^{2} \over 4}+\alpha_{m}^{2}
     g^{2}n}.
     \end{equation}
     So, the matrix elements of the evolution operator are:
     $$<n,J_{b},m \vert \exp{(\imath {\hat \Omega} t)} \vert
     n_{1},J_{b},m_{1}> =$$
     \begin{equation}
\delta_{n,n_{1}} \delta_{m,m_{1}}
     \left\{\cos{(\Omega_{n+1,m} t)} + {\imath \delta \over
     2\Omega_{n+1,m}}\sin{(\Omega_{n+1,m} t)}\right\}~,
     \end{equation}

     $$<n,J_{c},m \vert \exp{(\imath {\hat \Omega} t)} \vert
     n_{1},J_{c},m_{1}> = $$
     \begin{equation}
\delta_{n,n_{1}} \delta_{m,m_{1}}
     \left\{\cos{(\Omega_{n,m} t)} - {\imath \delta \over
     2\Omega_{n,m}}\sin{(\Omega_{n,m} t)}\right\}~,
     \end{equation}

     $$<n,J_{b},m \vert \exp{(\imath {\hat \Omega} t)} \vert
     n_{1},J_{c},m_{1}> =$$
     \begin{equation}
- \delta_{n+1,n_{1}} \delta_{m,m_{1}}
     g \alpha_{m}\sqrt{n+1} \cdot {\sin{(\Omega_{n+1,m} t)} \over
     \Omega_{n+1,m}}~.
     \end{equation}

     In the experiment [2] the average total population
$$n_{b} = Tr\left\{{\hat n}_{b} \exp{(\imath {\hat \Omega} T)}
\rho_{0} \exp{(-\imath {\hat \Omega} T)}\right\}$$
of the  upper resonant level~$b$~after the atom passes through
the resonant cavity,  where T is the time of interaction,  was
detected. As follows from (4)-(6),

     \begin{equation}
     n_{b} = \sum_{n,m} f_{nn} n^{b}_{mm}
     \left\{\cos^{2}{(\Omega_{n+1,m} T)} + {\delta^{2} \over
     4 \Omega^{2}_{n+1,m}}            \sin^{2}{(\Omega_{n+1,m}
     T)}\right\}~,
     \end{equation}
     where the  atomic  and  field  subsystems  at the initial
instant of  time,  when  the  atom  enters  the  cavity,   are
independent and the initial density matrix of the system is
     $${\hat \rho}_{0} ={\hat \rho}^{A}_{0}
      \cdot {\hat \rho}^{F}_{0}~,$$
     while
     $${\hat \rho}^{A}_{0} = \sum_{m,m_{1}} n^{b}_{mm_{1}}\cdot
     \vert J_{b},m><J_{b},m_{1}\vert~,$$
     $${\hat \rho}^{F}_{0} = \sum_{n,n_{1}} f_{nn_{1}}\cdot
     \vert n><n_{1}\vert~.$$

     The cavity temperature  in  [2]  was  low,  so  that  the
initial field may be considered to be in its vacuum state:
     $$f_{n,n_{1}} = \delta_{n,0}\delta_{n_{1},0}.$$
     Then, in  case of exact resonance~$\delta=0$~the equation
(7) simplifies to
     \begin{equation}
     n_{b} = \sum_{m} n^{b}_{mm}
     \cos^{2}{(\theta_{m})}~,~\theta_{m}=\alpha_{m}gT~.
     \end{equation}
     Here~$n^{b}_{mm}$~is the initial population of the Zeeman
sublevel~$m$~of the upper level~$b$~. The resonant levels~$b$~
and~$c$~in the  experiment  [2] were the Rydberg states of the
rubidium atom with the angular momenta~$J_{b}=3/2$~and
~$J_{c}=3/2$~or~$J_{c}=5/2$~. The  upper level~$b$~was excited
from the ground state~$a$~with the angular momentum~$J_{a}=1/2$~
     by the  linearly-polarized laser pulse.  The evolution of
the atomic density matrix under the action of  the  excitation
pulse in the rotating-wave approximation is desribed by the equation
     \begin{equation}
     {d {\hat \rho}^{A} \over dt} = {\imath \over \hbar}
\left[ {\hat \rho}^{A},{\hat V}_{e}\right]~,
     \end{equation}
     where
     $${\hat V}_{e} = -({\hat {\bf d\/}}^{+}_{e} {\bf e\/}_{e}+
     {\hat {\bf d\/}}_{e} {\bf e\/}^{*}_{e})$$
     is the  interaction  operator of an atom with the cohernt
resonant laser   field,
  $${\hat {\bf d\/}}_{e}=\sum_{m_{b},m_{a}}
({\bf d\/}_{e})^{J_{a}J_{b}}_{m_{a}m_{b}}\cdot
   \vert J_{a},m_{a}><J_{b},m_{b}\vert,$$
     is the dipole moment operator of the atomic transition
     ~${J_{b} \to J_{a}},$
${\bf e\/}_{e}= e_{e}{\bf l\/} $~is the  slowly-varying
amplitude of   laser   field,~${\bf  l\/}  $~is  its  unit
polarization vector,  which constitutes the  angle~$\psi$~with
the polarization of the cavity field mode:
     $$l_{q}=\cos{\psi}\delta_{q,0}+{1                   \over
\sqrt{2}}\sin{\psi}(\delta_{q,-1}-\delta_{q,1})~.$$
     For purposes of simplicity we shall consider the exciting
pulses with small areas
     \begin{equation}
     \theta_{e} =   {\vert   d_{e}   \vert   \over    \hbar}
\int_{0}^{T_{e}} e_{e}(t)dt~\ll 1~
     \end{equation}
    (though in  case  of  transition~${3/2  \to  1/2}$~in  the
experiment [1]  the  following  results  do  not depend on the
exciting pulse area),~$d_{e}=d(J_{b}J_{a})$~is a reduced matrix
element of    the    dipole    moment    operator    for   the
transition~${J_{b} \to J_{a}}$~,~$T_{e}$~is the exciting pulse
duration. Under  the  limitation  (10)  we obtain from (9) the
density matrix of an atom (renormalized to unity trace)
     \begin{equation}
     {\hat \rho}^{A}_{0} =
{({\hat  {\bf  d\/}}^{+}_{e}  {\bf
l\/}){\hat \rho}^{A}_{in}({\hat {\bf d\/}}_{e} {\bf l\/})
\over
 Tr\left\{({\hat  {\bf  d\/}}^{+}_{e}  {\bf l\/}){\hat \rho}^{A}_{in}
({\hat {\bf d\/}}_{e} {\bf l\/})\right\}}
     \end{equation}
     at an instant when it enters the cavity. Here
$${\hat \rho}^{A}_{in}= {1 \over (2J_{a}+1)}\sum_{m}
     \vert J_{a},m><J_{a},m \vert$$
     is the initial equilibrium atomic density  matrix  before
the incidence of the exciting pulse.  As follows from (11) the
Zeeman sublevel populations in (8) are
     $$n^{b}_{mm} =   <J_{b},m  \vert{\hat  \rho}^{A}_{0}\vert
J_{b},m> = a_{m}\cos^{2}{\psi} + b_{m}\sin^{2}{\psi}~,$$
     where
   $$a_{m}=3\left(\matrix{J_{b}&1&J_{a} \cr -m&0&m}\right)^{2}~,$$
   $$b_{m}={3 \over 2} \left\{
\left(\matrix{J_{b}&1&J_{a} \cr -m&-1&m+1}\right)^{2}+
   \left(\matrix{J_{b}&1&J_{a}                             \cr
-m&1&m-1}\right)^{2}\right\}~.$$
     In case of transitions~${J_{b}=3/2 \to J_{a}=1/2}$~
     $$n^{b}_{-1/2,-1/2} = n^{b}_{1/2,1/2} = {1 \over 2} -
     {3 \over 8} \sin^{2}{\psi}~,$$
     $$n^{b}_{-3/2,-3/2} = n^{b}_{3/2,3/2} =
     {3 \over 8} \sin^{2}{\psi}~,$$
     and the total population (8) of the upper level after the
atom leaves the cavity is
     $$n_{b} =                \left(1-{3                 \over
4}\sin^{2}{(\psi)}\right)\cos^{2}{(\theta)} +
     {3 \over 4}\sin^{2}{(\psi)}\cos^{2}{(3\theta)}~,$$
     $$\theta = {gT \over 2\sqrt{15}}~,$$
     for the transitions~${J_{b}=3/2 \to J_{c}=3/2}$~and
     $$n_{b} =                \left(1-{3                 \over
4}\sin^{2}{(\psi)}\right)\cos^{2}{(\theta)} +
     {3 \over 4}\sin^{2}{(\psi)}\cos^{2}{\left(\sqrt{{3
\over 2}}\theta\right)}~,$$
     $$\theta = {gT \over \sqrt{10}}~,$$
     for the transitions~${J_{b}=3/2 \to J_{c}=5/2}$~.

     The atom  behaves  like  a  two-level  system  -  the
population
     $$n_{b} = \cos^{2}{(\theta)}$$
     oscillates with a single Rabi frequency - only in case
when the polarizations of the exciting laser pulse and  of the
cavity field   mode  coincide  -\\
~$\psi  =  0~,$~otherwise  the
oscillations with more than one Rabi frequencies appear.

     So,  the  Jaynes-Cummings  model generalized to
the case of the atomic levels degenerate in the projections of
the angular momenta on the quantization axis is a useful tools
for the description of the polarization properties of one-atom
micromasers.\\[3cm]

     {\bf References\/}\\[5mm]

     [1] Jaynes E T, Cummings F W 1963 {\it Proc. IEEE\/} {\bf
51 \/} 89

     [2] Walther H 1995 {\it Ann.N.Y.Acad.Sci.\/}  {\bf  755\/}
133

     [3] Sobelman I I 1972 {\it Introduction to the Theory of Atomic
     Spectra\/}

     ~~~~(New York:Pergamon)

\end{document}